\documentclass[pdftex,twocolumn,10pt,letterpaper]{extarticle}

\newcommand{\AUTHORS}{Authors here}
\newcommand{\TITLE}{Datacenter RPCs can be General and Fast}
\newcommand{\KEYWORDS}{Put your keywords here}
\newcommand{\CONFERENCE}{Somewhere}
\newcommand{\PAGENUMBERS}{no}       
\newcommand{\COLOR}{yes}
\newcommand{\showComments}{yes}
\newcommand{\comment}[1]{}
\newcommand{\onlyAbstract}{no}

\usepackage{ifthen}
\ifthenelse{\equal{\PAGENUMBERS}{yes}}{%
\usepackage[nohead,
            left=0.75in,right=0.75in,top=1in,
            footskip=0.5in,bottom=1in,     
            columnsep=0.33in
            ]{geometry}
}{%
\usepackage[noheadfoot,left=.75in,right=.75in,top=1in,
            footskip=0.5in,bottom=1in,
            columnsep=0.33in
	    ]{geometry}
}

\usepackage{epigraph}
\usepackage[font=small,labelfont={small,bf},aboveskip=4pt]{caption}
\usepackage{pifont}
\usepackage{appendix}
\usepackage[compact]{titlesec}              
\titlespacing{\paragraph}{0pt}{*1}{*1}      
\titlespacing*{\section}{0pt}{1.5ex plus 1ex minus .2ex}{1.3ex plus .2ex}

\usepackage[inline]{enumitem}
\setlist{itemsep=0pt,parsep=0pt,nosep}             

\usepackage{fancyhdr}

\ifthenelse{\equal{\PAGENUMBERS}{yes}}{%
  \pagestyle{plain}
}{%
  \pagestyle{empty}
}

\usepackage[numbers,sort]{natbib}

\usepackage{booktabs}
\usepackage[binary-units=true]{siunitx}
\usepackage{multirow}
\usepackage{color}
\usepackage{colortbl}
\usepackage{float}                           

\usepackage[T1]{fontenc}
\usepackage[utf8]{inputenc}
\usepackage{libertine}
\usepackage[libertine]{newtxmath}
\usepackage{inconsolata}
\usepackage{textcomp}
\usepackage[mathcal]{euscript}

\definecolor{commentgreen}{RGB}{2,112,10}
\usepackage{listings}
\newcommand{\V}[1]{\textit{#1}}
\usepackage{mathtools}
\usepackage{algorithm}
\usepackage[noend]{algpseudocode}

\ifthenelse{\equal{\COLOR}{yes}}{%
  \PassOptionsToPackage{hyphens}{url}\usepackage[colorlinks,citecolor=blue]{hyperref}
}{%
  \usepackage[pdfborder={0 0 0}]{hyperref}
}
\usepackage{url}

\usepackage{cleveref}
\crefformat{section}{\S#2#1#3} 
\crefformat{subsection}{\S#2#1#3}
\crefformat{subsubsection}{\S#2#1#3}

\hypersetup{%
pdfauthor = {\AUTHORS},
pdftitle = {\TITLE},
pdfsubject = {\CONFERENCE},
pdfkeywords = {\KEYWORDS},
bookmarksopen = {true}
}

\usepackage[font={small,it}]{subfig}
\definecolor{placeholderbg}{rgb}{0.85,0.85,0.85}

\usepackage[pdftex]{graphicx}
\usepackage{soul}
\usepackage{nicefrac}
\usepackage{setspace}	

\newcommand{\Rpc}{\texttt{Rpc} }
\newcommand{\Rpcs}{\mbox{\texttt{Rpc}'s }}
\newcommand{\ns}[1]{\SI{#1}{\nano\second}}
\newcommand{\us}[1]{\SI{#1}{\micro\second}}
\newcommand{\ms}[1]{\SI{#1}{\milli\second}}
\newcommand{\byte}[1]{\SI{#1}{\byte}}
\newcommand{\kbyte}[1]{\SI{#1}{\kilo\byte}}
\newcommand{\mbyte}[1]{\SI{#1}{\mega\byte}}

\newcommand{\Gbps}[1]{\SI{#1}{Gbps}}

\newcommand{\GbE}[1]{\SI{#1}{GbE}}

\newcommand{\Mrps}[1]{\SI{#1}{Mrps}}

\usepackage[absolute]{textpos}

\newfloat{acmcr}{b}{acmcr}

\newcommand{\note}[2]{
    \ifthenelse{\equal{\showComments}{yes}}{\textcolor{#1}{\small #2}}{}
}

\newcommand{\superscript}[1]{\ensuremath{^{\textrm{#1}}}}		
\def\intel{\superscript{\dag}}

\usepackage{titling}
\newcommand*{\TitleFont}{%
      \usefont{\encodingdefault}{\rmdefault}{b}{n}%
      \fontsize{16}{20}%
      \selectfont}

\date{}
\title{\vspace{-40pt}\textbf{\TitleFont \TITLE}}
\author{\large{{Anuj Kalia \quad Michael Kaminsky\intel \quad David G. Andersen}}\\
\normalsize{\textit{Carnegie Mellon University \quad \quad \intel Intel Labs}}\\
}

\usepackage{microtype}  

\begin{document}

\maketitle
\thispagestyle{empty}


\begin{abstract}
  
It is commonly believed that datacenter networking software must sacrifice
generality to attain high performance. The popularity of specialized
distributed systems designed specifically for niche technologies such as RDMA,
lossless networks, FPGAs, and programmable switches testifies to this belief.
In this paper, we show that such specialization is not necessary. eRPC is a new
general-purpose remote procedure call (RPC) library that offers performance
comparable to specialized systems, while running on commodity CPUs in
traditional datacenter networks based on either lossy Ethernet or lossless
fabrics. eRPC performs well in three key metrics: message rate for small
messages; bandwidth for large messages; and scalability to a large number of
nodes and CPU cores. It handles packet loss, congestion, and background request
execution. In microbenchmarks, one CPU core can handle up to 10 million small
RPCs per second, or send large messages at \Gbps{75}. We
port a production-grade implementation of Raft state machine replication to
eRPC without modifying the core Raft source code. We achieve \us{5.5} of
replication latency on lossy Ethernet, which is faster than or comparable to
specialized replication systems that use programmable switches, FPGAs, or RDMA.

\end{abstract}

\ifthenelse{\equal{\onlyAbstract}{no}}{%
\section{Introduction}
\label{sec:intro}

{\small\emph{``Using performance to justify placing functions in a low-level
subsystem must be done carefully. Sometimes, by examining the problem thoroughly,
the same or better performance can be achieved at the high level.''}}

\rightline{{\small --- End-to-end Arguments in System Design}}

Squeezing the best performance out of modern, high-speed datacenter networks
has meant painstaking specialization that breaks down the abstraction barriers
between software and hardware layers. The result has been an explosion of
co-designed distributed systems that depend on niche network technologies,
including RDMA~\cite{Mitchell:usenix2013, Dragojevic:nsdi2014,
Kalia:sigcomm2014, wang:socc2014, Dragojevic:sosp2015, Wei:sosp2015,
Poke:hpdc2015, Mitchell:usenix2016, Binnig:vldb2016, Zamanian:vldb2017},
lossless networks~\cite{Kalia:osdi2016, Liu:eurosys2017},
FPGAs~\cite{Istvan:nsdi2016, Istvan:vldb2017}, and programmable
switches~\cite{Jin:nsdi2018}. Add to that new distributed protocols with
incomplete specifications, the inability to reuse existing software, hacks to
enable consistent views of remote memory---and the typical developer is likely
to give up and just use kernel-based TCP\@.

These specialized technologies were deployed with the belief that placing their
functionality in the network will yield a large performance gain. In this
paper, we show that a general-purpose RPC library called eRPC can provide
state-of-the-art performance on commodity datacenter networks without
additional network support. This helps inform the debate about the
utility of additional in-network functionality vs purely end-to-end
solutions for datacenter applications.

eRPC provides three key performance features: high message rate for small
messages; high bandwidth for large messages; and scalability to a large number
of nodes and CPU cores. It handles packet loss, node failures, congestion
control, and long-running background requests. eRPC is \emph{not} an RDMA-based
system: it works well with only UDP packets over lossy Ethernet without
Priority Flow Control (PFC), although it also supports InfiniBand\@. Our goal
is to allow developers to use eRPC in unmodified systems. We use as test-cases
two existing systems: a production-grade implementation of
Raft~\cite{Ongaro:2014, www-willemt-raft} that is used in Intel's distributed
object store~\cite{www-intel-daos}, and Masstree~\cite{Mao2012:eurosys2012}. We
successfully integrate eRPC support with both without sacrificing performance.

The need for eRPC arises because the
communication software options available for datacenter networks leave much to
be desired. The existing options offer an undesirable trade-off between
performance and generality. Low-level interfaces such as
DPDK~\cite{www-dpdk-general} are fast, but lack features required by general
applications (e.g., DPDK provides only unreliable packet I/O\@.) On the other
hand, full-fledged networking stacks such as mTCP~\cite{Jeong:nsdi2014} leave
significant performance on the table. Absent networking options that provide
both high performance and generality, recent systems often choose to design and
implement their own communication layer using low-level
interfaces~\cite{Ongaro:sosp2011, Kalia:sigcomm2014, Dragojevic:nsdi2014,
Dragojevic:sosp2015, Poke:hpdc2015, Wei:sosp2015, Binnig:vldb2016,
Kalia:osdi2016}.

The goal of our work is to answer the question: Can a general-purpose
RPC library provide performance comparable to specialized systems? Our solution
is based on two key insights. First, we optimize for the common case, i.e.,
when messages are small~\cite{Atikoglu2012, Ousterhout:tocs2015}, the network
is congestion-free, and RPC handlers are short. Handling large messages,
congestion, and long-running RPC handlers requires expensive code paths, which
eRPC avoids whenever possible. Several eRPC components, including its API,
message format, and wire protocol are optimized for the common case. Second,
restricting each flow to at most one bandwidth-delay product (BDP) of
outstanding data effectively prevents packet loss caused by switch buffer
overflow for common traffic patterns. This is because datacenter switch buffers
are much larger than the network's BDP\@. For example, in our two-layer testbed
that resembles real deployments, each switch has \mbyte{12} of dynamic buffer,
while the BDP is only \kbyte{19}.

eRPC (\emph{efficient} RPC) is available at
\url{https://github.com/efficient/eRPC}. Our research contributions are:

\begin{enumerate}
\item We describe the design and implementation of a high-performance RPC
library for datacenter networks. This includes
\begin{enumerate*}[label={(\arabic*)}]
\item common-case optimizations that improve eRPC's performance for our target
workloads by up to 66\%;
\item techniques that enable zero-copy transmission in the presence of
retransmissions, node failures, and rate limiting; and
\item a scalable implementation whose NIC memory footprint is independent of the
number of nodes in the cluster.
\end{enumerate*}

\item We are the first to show experimentally that state-of-the-art networking
performance can be achieved without lossless fabrics. We show that eRPC
performs well in a 100-node cluster with lossy Ethernet without PFC\@. Our
microbenchmarks on two lossy Ethernet clusters show that eRPC can:
\begin{enumerate*}[label={(\arabic*)}]
\item provide \us{2.3} median RPC latency;
\item handle up to 10 million RPCs per second with one core;
\item transfer large messages at \Gbps{75} with one core;
\item maintain low switch queueing during incast; and
\item maintain peak performance with 20000 connections per node (two million
connections cluster-wide).
\end{enumerate*}

\item We show that eRPC can be used as a high-performance drop-in networking
library for existing software. Notably, we implement a replicated in-memory
key-value store with a production-grade version of Raft~\cite{www-willemt-raft,
Ongaro:2014} without modifying the Raft source code. Our three-way replication
latency on lossy Ethernet is \us{5.5}, which is competitive with existing
specialized systems that use programmable switches
(NetChain~\cite{Jin:nsdi2018}), FPGAs~\cite{Istvan:nsdi2016}, and RDMA
(DARE~\cite{Poke:hpdc2015}).

\end{enumerate}


\section{Background and motivation}
\label{sec:background}
We first discuss aspects of modern datacenter networks relevant to eRPC\@. Next,
we discuss limitations of existing networking software that underlie the need
for eRPC\@.

\subsection{High-speed datacenter networking}
Modern datacenter networks provide tens of Gbps per-port bandwidth and a few
microseconds round-trip latency~\cite[\S{}2.1]{Zhu:sigcomm2015}. They support
polling-based network I/O from userspace, eliminating interrupts and system
call overhead from the datapath~\cite{Dalton:nsdi2018, Firestone:nsdi2018}.
eRPC uses userspace networking with polling, as in most prior high-performance
networked systems~\cite{Dragojevic:nsdi2014, Ousterhout:tocs2015,
Kalia:osdi2016, Jin:nsdi2018}.

eRPC works well in commodity, lossy datacenter networks. We found that
restricting each flow to one BDP of outstanding data prevents most packet
drops even on lossy networks. We discuss these aspects below.

\paragraph{Lossless fabrics.} Lossless packet delivery is a link-level feature
that prevents congestion-based packet drops. For example, PFC for Ethernet
prevents a link's sender from overflowing the receiver's buffer by using pause
frames. Some datacenter operators, including Microsoft, have deployed PFC at
scale. This was done primarily to support RDMA, since existing RDMA NICs
perform poorly in the presence of packet loss~\cite[\S1]{Zhu:sigcomm2015}.
Lossless fabrics are useful even without RDMA: Some systems that do not use
remote CPU bypass leverage losslessness to avoid the complexity and overhead of
handling packet loss in software~\cite{Kalia:sigcomm2014, Kalia:osdi2016,
Liu:eurosys2017}.

Unfortunately, PFC comes with a host of problems, including head-of-line
blocking, deadlocks due to cyclic buffer dependencies, and complex switch
configuration;~\citet{Mittal:sigcomm2018} discuss these problems in detail. In
our experience, datacenter operators are often unwilling to deploy PFC due to
these problems. Using simulations, Mittal~et~al. show that a new RDMA NIC
architecture called IRN with improved packet loss handling can work well in
lossy networks. Our BDP flow control is inspired by their work; the differences
between eRPC's and IRN's transport are discussed in
Section~\ref{subsubsec:cc_irn}. Note that, unlike IRN, eRPC is a real system,
and it does not require RDMA NIC support.

\paragraph{Switch buffer $\gg$ BDP\@.} The increase in datacenter bandwidth has
been accompanied by a corresponding decrease in round-trip time (RTT),
resulting in a small BDP\@.  Switch buffers have
grown in size, to the point where ``shallow-buffered'' switches that use SRAM
for buffering now provide tens of megabytes of shared buffer. Much of this
buffer is dynamic, i.e., it can be dedicated to an
incast's target port, preventing packet drops from buffer overflow. For
example, in our two-layer \GbE{25} testbed that resembles real datacenters
(Table~\ref{tab:clusters}), the RTT between two nodes connected to different
top-of-rack (ToR) switches is \us{6}, so the BDP is \kbyte{19}. This is
unsurprising: for example, the BDP of the two-tier \GbE{10} datacenter used in
pFabric is \kbyte{18}~\cite{Alizadeh:sigcomm2013}.

In contrast to the small BDP, the Mellanox Spectrum switches in our cluster
have \mbyte{12} in their dynamic buffer pool~\cite{www-warner-buffer}.
Therefore, the switch can ideally tolerate a 640-way incast. The popular
Broadcom Trident-II chip used in datacenters at Microsoft and Facebook has a
\mbyte{9} dynamic buffer~\cite{Zhu:sigcomm2015,
www-facebook-fboss}.~\citet{Zhang:jsac2014} have made a similar observation
(i.e., buffer $\gg$ BDP) for gigabit Ethernet.

In practice, we wish to support approximately 50-way incasts: congestion
control protocols deployed in real datacenters are tested against comparable
incast degrees. For example, DCQCN and Timely use up to 20- and 40-way incasts,
respectively~\cite{Zhu:sigcomm2015, Mittal:sigcomm2015}. This is much smaller
than 640, allowing substantial tolerance to technology variations, i.e., we
expect the switch buffer to be large enough to prevent most packet drops in
datacenters with different BDPs and switch buffer sizes. Nevertheless, it is
unlikely that the BDP-to-buffer ratio will grow substantially in the near
future: newer \GbE{100} switches have even larger buffers (\mbyte{42} in
Mellanox's Spectrum-2 and \mbyte{32} in Broadcom's Trident-III), and NIC-added
latency is continuously decreasing. For example, we measured InfiniBand's RTT
between nodes under different ToR's to be only \us{3.1}, and Ethernet has
historically caught up with InfiniBand's performance.

\subsection{Limitations of existing options}
Two reasons underlie our choice to design a new general-purpose RPC system for
datacenter networks:  First, existing datacenter networking software options sacrifice
performance or generality, preventing unmodified applications from using the
network efficiently. Second, co-designing storage software with the network is
increasingly popular, and is largely seen as necessary to achieve maximum
performance. However, such specialization has well-known drawbacks, which can
be avoided with a general-purpose communication layer that also provides high
performance. We describe a representative set of currently available options
and their limitations below, roughly in order of increasing performance and
decreasing generality.

Fully-general networking stacks such as mTCP~\cite{Jeong:nsdi2014} and
IX~\cite{Belay:osdi2014} allow legacy sockets-based applications to run
unmodified. Unfortunately, they leave substantial performance on the table,
especially for small messages. For example, one server core can handle around
1.5 million and 10 million \byte{64} RPC requests per second with
IX~\cite{Belay:osdi2014} and eRPC, respectively.

Some recent RPC systems can perform better, but are designed for specific use
cases. For example, RAMCloud RPCs~\cite{Ousterhout:tocs2015} are designed for
low latency, but not high throughput. In RAMCloud, a single dispatch thread
handles all network I/O, and request processing is done by other worker
threads. This requires inter-thread communication for every request, and limits
the system's network throughput to one core. FaRM
RPCs~\cite{Dragojevic:nsdi2014} use RDMA writes over connection-based hardware
transports, which limits scalability and prevents use in non-RDMA environments.

Like eRPC, our prior work on FaSST RPCs~\cite{Kalia:osdi2016} uses only
datagram packet I/O, but requires a lossless fabric. FaSST RPCs do not handle
packet loss, large messages, congestion, long-running request handlers, or node
failure; researchers have believed that supporting these features in
software (instead of NIC hardware) would substantially degrade
performance~\cite{Dragojevic:deb2017}. We show that with careful design, we can
support all these features and still match FaSST's performance, while running
on a lossy network. This upends conventional wisdom that losslessness or
NIC support is necessary for high performance.

\subsection{Drawbacks of specialization}
Co-designing distributed systems with network hardware is a well-known
technique to improve performance. Co-design with RDMA is popular, with numerous
examples from key-value stores~\cite{Mitchell:usenix2013, Kalia:sigcomm2014,
Dragojevic:nsdi2014, Wei:sosp2015, Wang:sc2015}, state machine
replication~\cite{Poke:hpdc2015}, and transaction processing
systems~\cite{Dragojevic:sosp2015, Wei:sosp2015, Chen:eurosys2016,
Kim:sigcomm2018}. Programmable switches allow in-network optimizations such as
reducing network round trips for distributed protocols~\cite{Li:osdi2016,
Jin:nsdi2018, Li:sosp2017}, and in-network caching~\cite{Jin:sosp2017}.
Co-design with FPGAs is an emerging technique~\cite{Istvan:nsdi2016}.

While there are advantages of co-design, such specialized systems are
unfortunately very difficult to design, implement, and deploy. Specialization
breaks abstraction boundaries between components, which prevents reuse of
components and increases software complexity. Building distributed storage
systems requires tremendous programmer effort, and co-design typically mandates
starting from scratch, with new data structures, consensus protocols, or
transaction protocols. Co-designed systems often cannot reuse existing
codebases or protocols, tests, formal specifications, programmer hours, and
feature sets. Co-design also imposes deployment challenges beyond needing
custom hardware: for example, using programmable switches requires user control
over shared network switches, which may not be allowed by datacenter operators;
and, RDMA-based systems are unusable with current NICs in datacenters that do
not support PFC.

In several cases, specialization does not provide even a performance advantage.
Our prior work shows that RPCs outperform RDMA-based designs for applications
like key-value stores and distributed transactions, with the same amount of
CPU~\cite{Kalia:sigcomm2014, Kalia:osdi2016}. This is primarily because
operations in these systems often require multiple remote memory accesses that
can be done with one RPC, but require multiple RDMAs. In this paper
(\S~\ref{subsec:raft}), we show that RPCs perform comparably with switch- and
FPGA-based systems for replication, too.


\section{\lowercase{e}RPC overview}
\label{sec:overview}


We provide an overview of eRPC's API and threading model below. In these
aspects, eRPC is similar to existing high-performance RPC systems like
Mellanox's Accelio~\cite{www-mellanox-accelio} and FaRM. eRPC's threading model
differs in how we sometimes run long-running RPC handlers in ``worker''
threads (\S~\ref{subsec:worker_threads}).

eRPC implements RPCs on top of a transport layer that provides basic unreliable
packet I/O, such as UDP or InfiniBand's Unreliable Datagram transport. A
userspace NIC driver is required for good performance. Our primary
contribution is the design and implementation of end-host mechanisms and a
network transport (e.g., wire protocol and congestion control) for the
commonly-used RPC API.

\subsection{RPC API}

RPCs execute at most once, and are asynchronous to avoid stalling on network
round trips; intra-thread concurrency is provided using an event loop. RPC
servers register request handler functions with unique request types; clients
use these request types when issuing RPCs, and get continuation callbacks on
RPC completion. Users store RPC messages in opaque, DMA-capable buffers
provided by eRPC, called msgbufs; a library that provides marshalling and
unmarshalling can be used as a layer on top of eRPC\@.

Each user thread that sends or receives RPCs creates an exclusive \Rpc endpoint
(a C\texttt{++} object). Each \Rpc endpoint contains an RX and TX queue for
packet I/O, an event loop, and several \emph{sessions}. A session is a
one-to-one connection between two \Rpc endpoints, i.e., two user threads. The
client endpoint of a session is used to send requests to the user thread at the
other end. A user thread may participate in multiple sessions, possibly playing
different roles (i.e., client or server) in different sessions.

User threads act as ``dispatch'' threads: they must periodically run their \Rpc
endpoint's event loop to make progress. The event loop performs the bulk of
eRPC's work, including packet I/O, congestion control, and management
functions. It invokes request handlers and continuations, and dispatches
long-running request handlers to worker threads
(\S~\ref{subsec:worker_threads}).

\paragraph{Client control flow:} \texttt{rpc->enqueue\_request()} queues a
request msgbuf on a session, which is transmitted when the user runs
\texttt{rpc}'s event loop.  On receiving the response, the event loop copies it
to the client's response msgbuf and invokes the continuation callback.

\paragraph{Server control flow:} The event loop of the \texttt{Rpc} that owns
the server session invokes (or dispatches) a request handler on receiving a
request. We allow \emph{nested} RPCs, i.e., the handler need not enqueue a
response before returning. It may issue its own RPCs and call
\texttt{enqueue\_response()} for the first request later when all dependencies
complete.

\subsection{Worker threads}
\label{subsec:worker_threads}
A key design decision for an RPC system is which thread runs an RPC handler.
Some RPC systems such as RAMCloud use dispatch threads for only network I/O.
RAMCloud's dispatch threads communicate with \emph{worker} threads that run
request handlers. At datacenter network speeds, however, inter-thread
communication is expensive: it reduces throughput and adds up to \ns{400} to
request latency~\cite{Ousterhout:tocs2015}. Other RPC systems such as Accelio
and FaRM avoid this overhead by running all request handlers directly in
dispatch threads~\cite{Dragojevic:nsdi2014, Kalia:sigcomm2014}. This latter
approach suffers from two drawbacks when executing long request handlers:
First, such handlers block other dispatch processing, increasing tail latency.
Second, they prevent rapid server-to-client congestion feedback, since the
server might not send packets while running user code.

Striking a balance, eRPC allows running request handlers in both dispatch
threads and worker threads: When registering a request handler, the programmer
specifies whether the handler should run in a dispatch thread. This is the only
additional user input required in eRPC. In typical use cases, handlers that
require up to a few hundred nanoseconds use dispatch threads, and longer
handlers use worker threads.

\subsection{Evaluation clusters}
\begin{table*}
\begin{center}
\small
\begin{tabular}{llllll}
\textbf{Name} & \textbf{Nodes} & \textbf{Network type} & \textbf{Mellanox NIC} & \textbf{Switches} & \textbf{Intel Xeon E5 CPU code} \\
\midrule
CX3 & 11 & InfiniBand & \Gbps{56} ConnectX-3 & One SX6036 & 2650 (8 cores) \\
CX4 & 100 & Lossy Ethernet & \Gbps{25} ConnectX-4 Lx & 5x SN2410, 1x SN2100 & 2640 v4 (10 cores) \\
CX5 & 8 & Lossy Ethernet & Dual-port \Gbps{40} ConnectX-5 & One SX1036 & 2697 v3 (14 c) or 2683 v4 (16 c)\\
\end{tabular}
\caption{Measurement clusters. CX4 and CX3 are CloudLab~\cite{Ricci:login14}
         and Emulab~\cite{White+:osdi2002} clusters, respectively.}
\label{tab:clusters}
\end{center}
\end{table*}

Table~\ref{tab:clusters} shows the clusters used in this paper. They include
two types of networks (lossy Ethernet, and lossless InfiniBand), and three
generations of NICs released between 2011 (CX3) and 2017 (CX5). eRPC works well
on all three clusters, showing that our design is robust to NIC and network
technology changes. We use traditional UDP on the Ethernet clusters (i.e., we
do not use RoCE), and InfiniBand's Unreliable Datagram transport on the
InfiniBand cluster.

Currently, eRPC is primarily optimized for Mellanox NICs. eRPC also works with
DPDK-capable NICs that support flow steering. For Mellanox Ethernet NICs, we
generate UDP packets directly with \texttt{libibverbs} instead of going through
DPDK, which internally uses \texttt{libibverbs} for these NICs.

Our evaluation primarily uses the large CX4 cluster, which resembles real-world
datacenters. The ConnectX-4 NICs used in CX4 are widely deployed in datacenters
at Microsoft and Facebook~\cite{Zhu:sigcomm2015, www-facebook-yosemite}, and
its Mellanox Spectrum switches perform similarly to Broadcom's Trident switches
used in these datacenters (i.e., both switches provide dynamic buffering,
cut-through switching, and less than \ns{500} port-to-port latency.) We use 100
nodes out of the 200 nodes in the shared CloudLab cluster. The six switches in
the CX4 cluster are organized as five ToRs with 40 \GbE{25} downlinks and five
\GbE{100} uplinks, for a 2:1 oversubscription. 

\section{\lowercase{e}RPC design}
\label{sec:design}
Achieving eRPC's performance goals requires careful design and implementation.
We discuss three aspects of eRPC's design in this section: scalability of our
networking primitives, the challenges involved in supporting zero-copy, and the
design of sessions. The next section discusses eRPC's wire protocol and
congestion control.  A recurring theme in eRPC's design is that we optimize for
the common case, i.e., when request handlers run in dispatch threads, RPCs are
small, and the network is congestion-free.

\subsection{Scalability considerations}
\label{subsec:scalability_considerations}
We chose plain packet I/O instead of RDMA writes~\cite{Dragojevic:nsdi2014,
Wei:sosp2015, Zamanian:vldb2017} to send messages in eRPC. This decision is
based on prior insights from our design of FaSST: First, packet I/O provides
completion queues that can scalably detect received packets.  Second, RDMA
caches connection state in NICs, which does not scale to
large clusters. We next discuss \emph{new} observations about NIC hardware
trends that support this design.

\subsubsection{Packet I/O scales well}
\label{subsubsec:packet_io_scales}
RPC systems that use RDMA writes have a \emph{fundamental} scalability
limitation. In these systems, clients write requests directly to per-client
circular buffers in the server's memory; the server must poll these buffers to
detect new requests.  The number of circular buffers grows with the number of
clients, limiting scalability.

With traditional userspace packet I/O, the NIC writes an incoming packet's
payload to a buffer specified by a descriptor pre-posted to the NIC's RX queue
(RQ) by the receiver host; the packet is dropped if the RQ is empty. Then, the
NIC writes an entry to the host's RX completion queue. The receiver host can
then check for received packets in constant time by examining the head of the
completion queue.

To avoid dropping packets due to an empty RQ with no descriptors, RQs must be
sized proportionally to the number of independent connected RPC endpoints
(\S~\ref{subsubsec:session_credits}). Older NICs experience cache thrashing
with large RQs, thus limiting scalability, but we find that newer NICs fare
better: While a Connect-IB NIC could support only 14 2K-entry RQs before
thrashing~\cite{Kalia:osdi2016}, we find that ConnectX-5 NICs do not thrash
even with 28 64K-entry RQs. This improvement is due to more intelligent
prefetching and caching of RQ descriptors, instead of a massive 64x increase in
NIC cache.

We use features of current NICs (e.g., multi-packet RQ descriptors that
identify several contiguous packet buffers) in novel ways  to guarantee a
\emph{constant} NIC memory footprint per CPU core, i.e., it does not depend on
the number of nodes in the cluster. This result can simplify the design of
future NICs (e.g., RQ descriptor caching is unneeded), but its current value is
limited to performance improvements because current NICs support very large
RQs, and are perhaps overly complex as a result. We discuss this in detail in
Appendix~\ref{sec:erpc_nic_footprint}.

\subsubsection{Scalability limits of RDMA}
\label{subsubsec:rdma}
\begin{figure}
  \centering
	\includegraphics[width=0.4\textwidth]{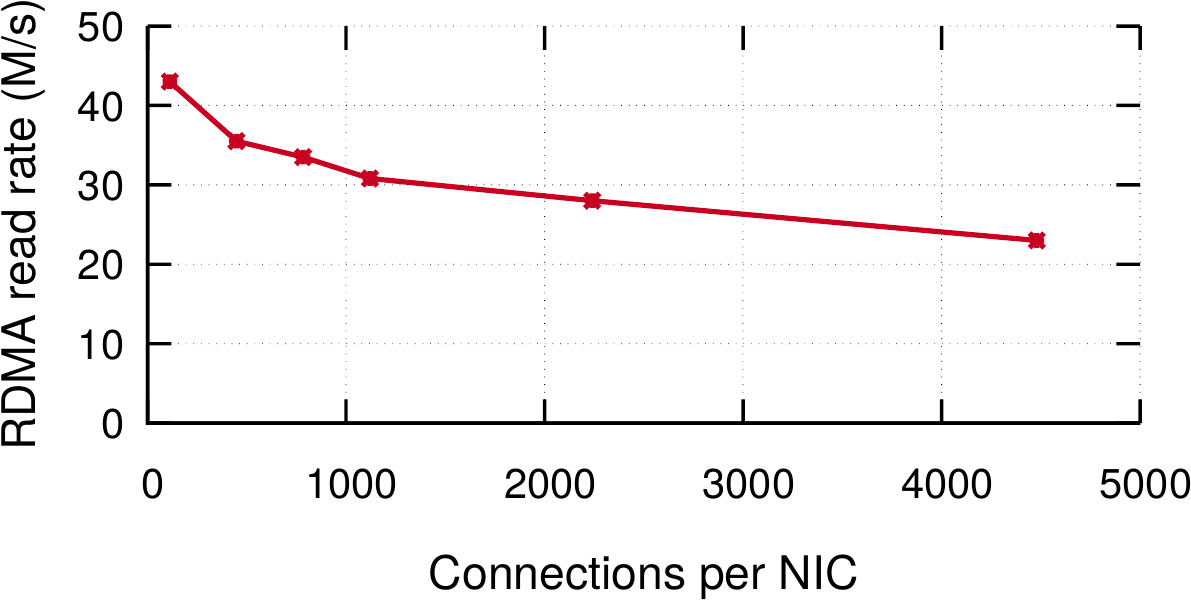}
	\caption{Connection scalability of ConnectX-5 NICs}
	\label{fig:qp_scalability}
\end{figure}

RDMA requires NIC-managed connection state. This limits scalability because
NICs have limited SRAM to cache connection state. The number of in-NIC
connections may be reduced by sharing them among CPU cores, but doing so
reduces performance by up to 80\%~\cite{Kalia:osdi2016}. 

Some researchers have hypothesized that improvements in NIC hardware will allow
using connected transports at large scale~\cite{Dragojevic:deb2017,
Zamanian:vldb2017}. To show that this is unlikely, we measure the connection
scalability of state-of-the-art ConnectX-5 NICs, released in 2017.  We repeat
the connection scalability experiment from FaSST, which was used to evaluate
the older Connect-IB NICs from 2012. We enable PFC on CX5 for this experiment
since it uses RDMA; PFC is disabled in all experiments that use eRPC. In the experiment,
each node creates a tunable number of connections to other nodes and issues
16-byte RDMA reads on randomly-chosen connections.
Figure~\ref{fig:qp_scalability} shows that as the number of connections
increases, RDMA throughput decreases, losing $\approx$50\% throughput with 5000
connections. This happens because NICs can cache only a few connections, and
cache misses require expensive DMA reads~\cite{Dragojevic:nsdi2014}. In
contrast, eRPC maintains its peak throughput with 20000 connections
(\S~\ref{subsec:session_scalability}).

ConnectX-5's connection scalability is, surprisingly, not substantially better
than Connect-IB despite the five-year advancement. A simple calculation shows
why this is hard to improve: In Mellanox's implementation, each connection
requires $\approx$\byte{375} of in-NIC connection state, and the NICs have
$\approx$\mbyte{2} of SRAM to store connection state as well as other data
structures and buffers~\cite{MellanoxPriv}. 5000 connections require
\mbyte{1.8}, so cache misses are unavoidable.

NIC vendors have been trying to improve RDMA's scalability for a
decade~\cite{Koop:cluster2008, crupnicoff2011dynamically}. Unfortunately, these
techniques do not map well to RPC workloads~\cite{Kalia:osdi2016}. Vendors have
not put more memory in NICs, probably because of cost and power overheads, and
market factors. The scalability issue of RDMA is exacerbated by the popularity
of \emph{multihost} NICs, which allow sharing a powerful NIC among 2--4
CPUs~\cite{www-facebook-yosemite, www-ornl-summit}.

eRPC replaces NIC-managed connection state with CPU-managed connection state.
This is an explicit design choice, based upon fundamental differences between
the CPU and NIC architectures.  NICs and CPUs will both cache recently-used
connection state.  CPU cache misses are served from DRAM, whereas NIC cache
misses are served from the CPU's memory subsystem over the slow PCIe bus. The
CPU's miss penalty is therefore much lower.  Second, CPUs have substantially
larger caches than the $\sim$\mbyte{2} available on a modern NIC, so the cache
miss \emph{frequency} is also lower.

\subsection{Challenges in zero-copy transmission}
\label{subsec:zero_copy}
\begin{figure}
  \centering
	\includegraphics[width=0.45\textwidth]{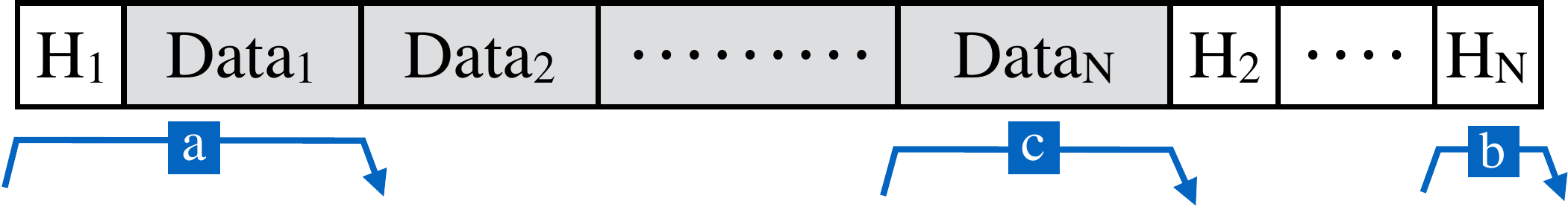}
	\caption{Layout of packet headers and data for an $N$-packet msgbuf. Blue
  arrows show NIC DMAs; the letters show the order in which the DMAs are
  performed for packets 1 and $N$\@.}
	\label{fig:msgbuf}
\end{figure}

eRPC uses zero-copy packet I/O to provide performance comparable to low-level
interfaces such as DPDK and RDMA. This section describes the challenges
involved in doing so.

\subsubsection{Message buffer layout}
eRPC provides DMA-capable message buffers to applications for zero-copy
transfers. A msgbuf holds one, possibly multi-packet message. It consists of
per-packet headers and data, arranged in a fashion optimized for small
single-packet messages (Figure~\ref{fig:msgbuf}). Each eRPC packet has a
header that contains the transport header, and eRPC metadata such as the request
handler type and sequence numbers. We designed a
msgbuf layout that satisfies two requirements.

\begin{enumerate}
\item The data region is contiguous to allow its use in applications as an
opaque buffer.
\item The first packet's data and header are contiguous. This allows the NIC to
fetch small messages with one DMA read; using multiple DMAs for small messages
would substantially increase NIC processing and PCIe use, reducing message rate
by up to 20\%~\cite{Kalia:usenix2016}.
\end{enumerate}

For multi-packet messages, headers for subsequent packets are at the end of the
message: placing header~2 immediately after the first data packet would violate
our first requirement. Non-first packets require two DMAs (header and data);
this is reasonable because the overhead for DMA-reading small headers is
amortized over the large data DMA\@.

\subsubsection{Message buffer ownership}
\label{subsubsec:buffer_ownership}

Since eRPC transfers packets directly from application-owned msgbufs, msgbuf
references must never be used by eRPC after msgbuf ownership is returned
to the application. In this paper, we discuss msgbuf ownership issues for only
clients; the process is similar but simpler for the server, since eRPC's
servers are passive (\S~\ref{sec:wire_protocol}). At clients, we must ensure
the following invariant: \emph{no eRPC transmission queue contains a reference
to the request msgbuf when the response is processed.} Processing the response
includes invoking the continuation, which permits the application to reuse the
request msgbuf. In eRPC, a request reference may be queued in the NIC's hardware DMA
queue, or in our software rate limiter (\S~\ref{subsec:cc}).

This invariant is maintained trivially when there are no retransmissions or
node failures, since the request must exit all transmission queues before
the response is received. The following \textbf{example} demonstrates the
problem with retransmissions. Consider a client that falsely suspects packet
loss and retransmits its request. The server, however, received the first copy of the
request, and its response reaches the client before the retransmitted request
is dequeued. Before processing the response and invoking the continuation, we
must ensure that there are no queued references to the request msgbuf. We
discuss our solution for the NIC DMA queue next, and for the rate limiter in
Appendix~\ref{sec:rl_zero_copy}.

The conventional approach to ensure DMA completion is to use ``signaled''
packet transmission, in which the NIC writes completion entries to the TX
completion queue. Unfortunately, doing so reduces message rates by up to 25\%
by using more NIC and PCIe resources~\cite{Kalia:sigcomm2014}, so we use
unsignaled packet transmission in eRPC.

Our method of ensuring DMA completion with unsignaled transmission is in line
with our design philosophy: we choose to make the common case (no
retransmission) fast, at the expense of invoking a more-expensive mechanism to
handle the rare cases. We flush the TX DMA queue after queueing a retransmitted
packet, which blocks until all queued packets are DMA-ed. This ensures the
required invariant: when a response is processed, there are no references to
the request in the DMA queue. This flush is moderately expensive
($\approx$\us{2}), but it is called during rare retransmission or node failure
events, and it allows eRPC to retain the 25\% throughput increase from
unsignaled transmission.

During server node failures, eRPC invokes continuations with error codes,
which also yield request msgbuf ownership. It is possible, although extremely
unlikely, that server failure is suspected while a request (not necessarily a
retransmission) is in the DMA queue or the rate limiter. Handling node
failures requires similar care as discussed above, and is discussed in detail
in Appendix~\ref{sec:machine_failure}.

\subsubsection{Zero-copy request processing}
\label{subsubsec:zero_copy_req}
Zero-copy reception is harder than transmission: To provide a contiguous
request msgbuf to the request handler at the server, we must strip headers from
received packets, and copy only application data to the target msgbuf. However,
we were able to provide zero-copy reception for our common-case workload
consisting of single-packet requests and dispatch-mode request handlers as
follows. eRPC owns the packet buffers DMA-ed by the NIC until it re-adds the
descriptors for these packets back to the receive queue (i.e., the NIC cannot
modify the packet buffers for this period.) This ownership guarantee allows
running dispatch-mode handlers without copying the DMA-ed request packet to a
dynamically-allocated msgbuf. Doing so improves eRPC's message rate by up to
16\% (\S~\ref{subsec:small_rpc_rate}).

\subsection{Sessions}
\label{subsec:session_impl}

Each session maintains multiple outstanding requests to keep the network pipe
full. Concurrently requests on a session can complete \emph{out-of-order} with
respect to each other. This avoids blocking dispatch-mode RPCs behind a
long-running worker-mode RPC. We support a constant number of concurrent
requests (default = 8) per session; additional requests are transparently
queued by eRPC\@. This is inspired by how RDMA connections allow a constant
number of operations~\cite{www-ibv-modify-qp}. A session uses an array of
\emph{slots} to track RPC metadata for outstanding requests.

Slots in server-mode sessions have an MTU-size preallocated msgbuf for use by
request handlers that issue short responses. Using the preallocated msgbuf does
not require user input: eRPC chooses it automatically at run time by examining
the handler's desired response size. This optimization avoids the overhead of
dynamic memory allocation, and improves eRPC's message rate by up to 13\%
(\S~\ref{subsec:small_rpc_rate}).

\subsubsection{Session credits}
\label{subsubsec:session_credits}
eRPC limits the number of unacknowledged packets on a session for two reasons.
First, to avoid dropping packets due to an empty RQ with no descriptors, the
number of packets that may be sent to an \Rpc must not exceed the size of its
RQ ($|RQ|$). Because each session sends packets independently of others, we
first limit the number of sessions that an \Rpc can participate in. Each
session then uses \emph{session credits} to implement packet-level flow
control: we limit the number of packets that a client may send on a session
before receiving a reply, allowing the server \Rpc to replenish used RQ
descriptors before sending more packets.

Second, session credits automatically implement end-to-end flow control, which
reduces switch queueing (\S~\ref{subsec:cc}). Allowing $\V{BDP}/\V{MTU}$
credits per session ensures that each session can achieve line
rate.~\citet{Mittal:sigcomm2018} have proposed similar flow control for RDMA
NICs (\S~\ref{subsubsec:cc_irn}).

A client session starts with a quota of $C$ packets. Sending a
packet to the server consumes a credit, and receiving a packet replenishes a
credit. An \Rpc can therefore participate in up to $\nicefrac{|RQ|}{C}$
sessions, counting both server-mode and client-mode sessions; session creation
fails after this limit is reached. We plan to explore statistical multiplexing
in the future.

\subsubsection{Session scalability}
\label{subsubsec:session_scalability}
eRPC's scalability depends on the user's desired value of $C$, and the number
and size of RQs that the NIC and host can effectively support. Lowering $C$
increases scalability, but reduces session throughput by restricting the
session's packet window. Small values of $C$ (e.g., $C = 1$) should be used in
applications that (a) require only low latency and small messages, or (b) whose
threads participate in many sessions. Large values (e.g., $\V{BDP}/\V{MTU}$)
should be used by applications whose sessions individually require high
throughput.

Modern NICs can support several very large RQs, so NIC RQ capacity limits
scalability only on older NICs. In our evaluation, we show that eRPC can handle
20000 sessions with 32 credits per session on the widely-used ConnectX-4 NICs.
However, since each RQ entry requires allocating a packet buffer in host
memory, needlessly large RQs waste host memory and should be avoided.

\section{Wire protocol}
\label{sec:wire_protocol}

We designed a wire protocol for eRPC that is optimized for small RPCs and
accounts for per-session credit limits. For simplicity, we chose a simple
\emph{client-driven} protocol, meaning that each packet sent by the server is
in response to a client packet. A client-driven protocol has fewer ``moving
parts'' than a protocol in which both the server and client can independently
send packets. Only the client maintains wire protocol state that is rolled back
during retransmission. This removes the need for client-server coordination
before rollback, reducing complexity. A client-driven protocol also shifts the
overhead of rate limiting entirely to clients, freeing server CPU that is
often more valuable.

\subsection{Protocol messages}
\label{sec:protocol_messages}
\begin{figure}
  \centering
	\includegraphics[width=0.45\textwidth]{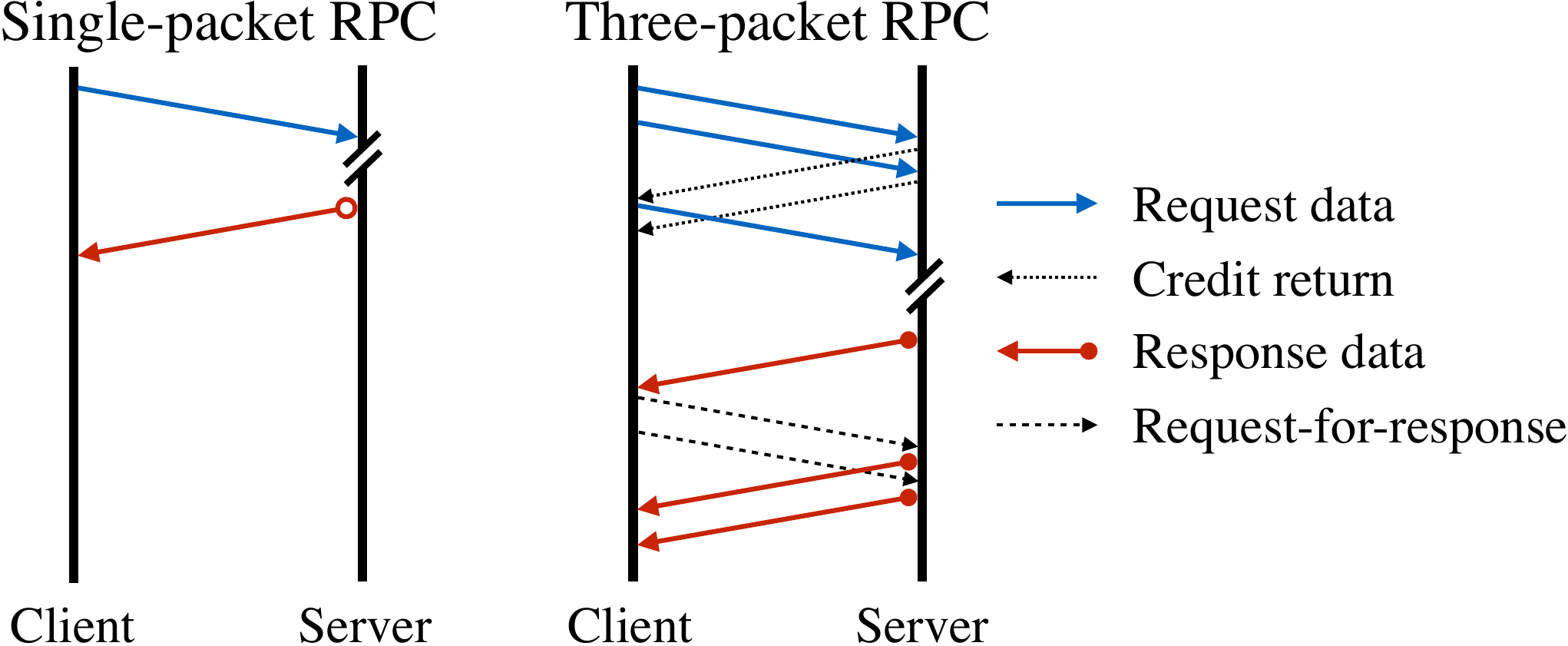}
	\caption{Examples of eRPC's wire protocol, with 2 credits/session.}
	\label{fig:protocol}
\end{figure}

Figure~\ref{fig:protocol} shows the packets sent with $C = 2$ for a small
single-packet RPC, and for an RPC whose request and response require three
packets each. Single-packet RPCs use the fewest packets possible. The client
begins by sending a window of up to $C$ request data packets. For each request
packet except the last, the server sends back an explicit \emph{credit return}
(CR) packet; the credit used by the last request packet is implicitly returned
by the first response packet.

Since the protocol is client-driven, the server cannot immediately send
response packets after the first. Subsequent response packets are triggered by
\emph{request-for-response} (RFR) packets that the client sends after receiving
the first response packet. This increases the latency of multi-packet responses
by up to one RTT\@. This is a fundamental drawback of client-driven protocols;
in practice, we found that the added latency is less than 20\% for responses
with four or more packets.

CRs and RFRs are tiny \byte{16} packets, and are sent only for large
multi-packet RPCs. The additional overhead of sending these tiny packets is
small with userspace networking that our protocol is designed for, so we do not
attempt complex optimizations such as cumulative CRs or RFRs. These
optimizations may be worthwhile for kernel-based networking stacks, where
sending a \byte{16} packet and an MTU-sized packet often have comparable CPU
cost.

\subsection{Congestion control}
\label{subsec:cc}
Congestion control for datacenter networks aims to reduce switch queueing,
thereby preventing packet drops and reducing RTT. Prior high-performance
RPC implementations such as FaSST do not implement congestion control, and some
researchers have hypothesized that doing so will substantially reduce
performance~\cite{Dragojevic:deb2017}. Can effective congestion control be
implemented efficiently in software? We show that optimizing for uncongested
networks, and recent advances in software rate limiting allow congestion
control with only 9\% overhead (\S~\ref{subsec:small_rpc_rate}).

\subsubsection{Available options}
\label{subsubsec:cc_options}
Congestion control for high-speed datacenter networks is an evolving area of
research, with two major approaches for commodity hardware: RTT-based
approaches such as Timely~\cite{Mittal:sigcomm2015}, and ECN-based approaches
such as DCQCN~\cite{Zhu:sigcomm2015}. Timely and DCQCN have been deployed at
Google and Microsoft, respectively. We wish to use these protocols since they
have been shown to work at scale.

Both Timely and DCQCN are rate-based: client use the congestion signals to
adjust per-session sending rates. We implement Carousel's rate
limiter~\cite{Saeed:sigcomm2017}, which is designed to efficiently handle a
large number of sessions. Carousel's design works well for us as-is, so we omit
the details.

eRPC includes the hooks and mechanisms to easily implement either Timely or
DCQCN\@. Unfortunately, we are unable to implement DCQCN because none of our
clusters performs ECN marking\footnote{The Ethernet switch in our private CX5
cluster does not support ECN marking~\cite[p.~839]{www-mellanox-mlnx-os}; we do
not have admin access to the shared CloudLab switches in the public CX4
cluster; and InfiniBand NICs in the CX3 cluster do not relay ECN marks to
software.}. Timely can be implemented entirely in software, which made it our
favored approach. eRPC runs all three Timely components---per-packet RTT
measurement, rate computation using the RTT measurements, and rate
limiting---at client session endpoints. For \Rpcs that host only server-mode
endpoints, there is no overhead due to congestion control.

\subsubsection{Common-case optimizations}
\label{subsubsec:cc_opt}
We use three optimizations for our common-case workloads. Our evaluation shows
that these optimizations reduce the overhead of congestion control from 20\% to
9\%, and that they do not reduce the effectiveness of congestion control. The
first two are based on the observation that datacenter networks are typically
uncongested. Recent studies of Facebook's datacenters support this
claim:~\citet{Roy:sigcomm2015} report that 99\% of all datacenter links are less
than 10\% utilized at one-minute timescales.~\citet[Fig. 6]{Zhang:imc2017}
report that for Web and Cache traffic, 90\% of top-of-rack switch links, which
are the most congested switches, are less than 10\% utilized at \us{25}
timescales.

When a session is uncongested, RTTs are low and Timely's computed rate for the
session stays at the link's maximum rate; we refer to such sessions as
\emph{uncongested}.

\begin{enumerate}
\item \textbf{Timely bypass.} If the RTT of a packet received on an uncongested
session is smaller than Timely's low threshold, below which it performs
additive increase, we do not perform a rate update. We use the recommended
value of \us{50} for the low threshold~\cite{Mittal:sigcomm2015, Zhu:conext2016}.

\item \textbf{Rate limiter bypass.} For uncongested sessions, we transmit
packets directly instead of placing them in the rate limiter.

\item \textbf{Batched timestamps for RTT measurement.} Calling \texttt{rdtsc()}
costs \ns{8} on our hardware, which is substantial when processing millions of
small packets per second. We reduce timer overhead by sampling it once per RX or
TX batch instead of once per packet.
\end{enumerate}

\subsubsection{Comparison with IRN}
\label{subsubsec:cc_irn}
IRN~\cite{Mittal:sigcomm2018} is a new RDMA NIC architecture designed for lossy
networks, with two key improvements. First, it uses BDP flow control to limit
the outstanding data per RDMA connection to one BDP. Second, it uses
efficient selective acks instead of simple go-back-N for packet loss recovery.

IRN was evaluated with simulated switches that have small (60--\kbyte{480})
static, per-port buffers. In this buffer-deficient setting, they found SACKs
necessary for good performance. However, dynamic-buffer switches are the
de-facto standard in current datacenters. As a result, packet losses are very
rare with only BDP flow control, so we currently do not implement SACKs,
primarily due to engineering complexity. eRPC's dependence on dynamic switch
buffers can be reduced by implementing SACK.

With small per-port switch buffers, IRN's maximum RTT is a few hundred microseconds,
allowing a $\sim$\us{300} retransmission timeout (RTO). However, the \mbyte{12}
dynamic buffer in our main CX4 cluster (\Gbps{25}) can add up to \ms{3.8} of
queueing delay. Therefore, we use a conservative \ms{5} RTO.

\subsection{Handling packet loss}
\label{subsec:packet_loss}
For simplicity, eRPC treats reordered packets as losses by dropping them. This
is not a major deficiency because datacenter networks typically use ECMP for
load balancing, which preserves intra-flow ordering~\cite{Zhou:eurosys2014,
Guo:sigcomm2015, Zhang:imc2017} except during rare route churn events. Note
that current RDMA NICs also drop reordered packets~\cite{Mittal:sigcomm2018}.

On suspecting a lost packet, the client rolls back the request's wire protocol
state using a simple go-back-N mechanism. It then reclaims credits used for the
rolled-back transmissions, and retransmits from the updated state. The server
never runs the request handler for a request twice, guaranteeing at-most-once
RPC semantics.

In case of a false positive, a client may violate
the credit agreement by having more packets outstanding to the server than its
credit limit. In the extremely rare case that such an erroneous loss detection
occurs \emph{and} the server's RQ is out of descriptors, eRPC will have
``induced'' a real packet loss. We allow this possibility and handle the
induced loss like a real packet loss. 

\section{Microbenchmarks}
\label{sec:microbenchmarks}

eRPC is implemented in 6200 SLOC of C\texttt{++}, excluding tests and
benchmarks. We use static polymorphism to create an \Rpc class that works with
multiple transport types without the overhead of virtual function calls. In
this section, we evaluate eRPC's latency, message rate, scalability, and
bandwidth using microbenchmarks. To understand eRPC's performance in commodity
datacenters, we primarily use the large CX4 cluster. We use CX5 and CX3 for
their more powerful NICs and low-latency InfiniBand, respectively. eRPC's
congestion control is enabled by default.

\subsection{Small RPC latency}
\label{subsec:small_rpc_latency}
\begin{table}
\centering
\small
\begin{tabular}{lccc}
\textbf{Cluster} & CX3 (InfiniBand) & CX4 (Eth) & CX5 (Eth) \\
\midrule
\textbf{RDMA read} & \us{1.7} & \us{2.9} & \us{2.0} \\
\textbf{eRPC} & \us{2.1} & \us{3.7} & \us{2.3} \\
\end{tabular}
\caption{Comparison of median latency with eRPC and RDMA}
\label{tab:small_rpc_latency}
\end{table}
How much latency does eRPC add? Table~\ref{tab:small_rpc_latency} compares the
median latency of \byte{32} RPCs and RDMA reads between two nodes connected to
the same ToR switch. Across all clusters, eRPC is at most \ns{800} slower than
RDMA reads. 

eRPC's median latency on CX5 is only \us{2.3}, showing that latency with
commodity Ethernet NICs and software networking is much lower than the
widely-believed value of 10--\us{100}~\cite{Jin:nsdi2018, Panda2016}. CX5's
switch adds \ns{300} to every layer-3 packet~\cite{www-mellanox-sx1036},
meaning that end-host networking adds only $\approx$\ns{850} each at the client
and server. This is comparable to switch-added latency. We discuss this further
in~\S~\ref{subsec:raft}.

\subsection{Small RPC rate}
\label{subsec:small_rpc_rate}
\begin{figure}
  \centering
	\includegraphics[width=0.45\textwidth]{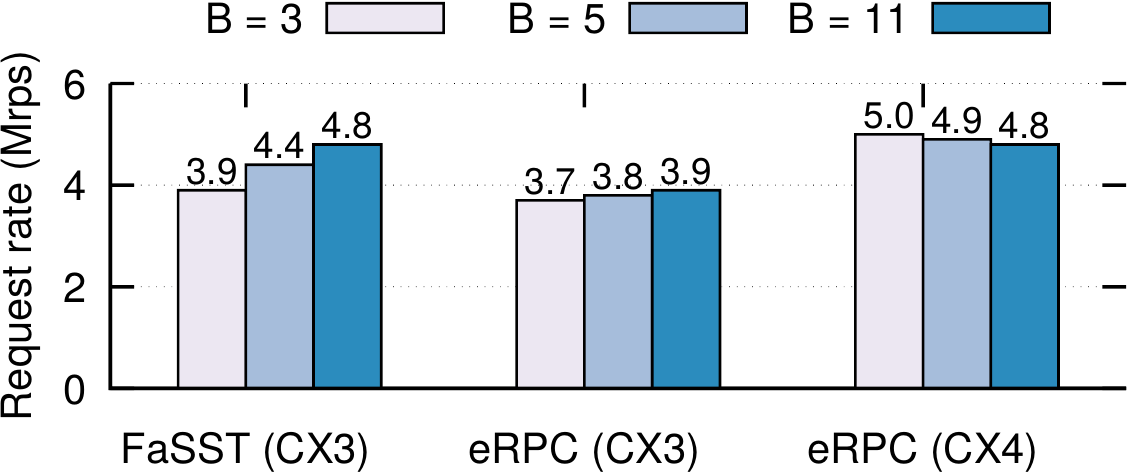}
	\caption{Single-core small-RPC rate with $B$ requests per batch}
	\label{fig:small_rpc_tput}
\end{figure}
What is the CPU cost of providing generality in an RPC system? We compare
eRPC's small message performance against FaSST RPCs, which outperform other RPC
systems such as FaRM~\cite{Kalia:osdi2016}. FaSST RPCs are \emph{specialized} for
single-packet RPCs in a lossless network, and they do not handle congestion.

We mimic FaSST's experiment setting: one thread per node in an 11-node cluster,
each of which acts each acts as both RPC server and client. Each thread issues
batches of $B$ requests, keeping multiple request batches in flight to hide
network latency. Each request in a batch is sent to a randomly-chosen remote
thread. Such batching is common in key-value stores and distributed online
transaction processing. Each thread keeps up to 60 requests in flight, spread
across all sessions. RPCs are \byte{32} in size. We compare eRPC's performance
on CX3 (InfiniBand) against FaSST's reported numbers on the same cluster. We
also present eRPC's performance on the CX4 Ethernet cluster. We omit CX5 since
it has only 8 nodes.

Figure~\ref{fig:small_rpc_tput} shows that eRPC's per-thread request issue rate is at
most 18\% lower than FaSST across all batch sizes, and only 5\% lower for $B =
3$. This performance drop is acceptable since eRPC is a full-fledged RPC
system, whereas FaSST is highly specialized. On CX4, each thread issues 5
million requests per second (Mrps) for $B = 3$; due to the experiment's symmetry,
it simultaneously also handles incoming requests from remote threads at \Mrps{5}.
Therefore, each thread processes 10 million RPCs per second.

\begin{table}
\begin{center}
\small
\setlength\tabcolsep{4pt}
\begin{tabular}{lll}
\textbf{Action} & \textbf{RPC rate} & \textbf{\% loss} \\
\midrule
Baseline (with congestion control) & 4.96 M/s & -- \\
\midrule
Disable batched RTT timestamps (\cref{subsec:cc}) & 4.84 M/s & 2.4\% \\
Disable Timely bypass (\cref{subsec:cc}) & 4.52 M/s & 6.6\% \\
Disable rate limiter bypass (\cref{subsec:cc}) & 4.30 M/s & 4.8\% \\
\midrule
Disable multi-packet RQ (\cref{subsubsec:packet_io_scales}) & 4.06 M/s & 5.6\% \\
Disable preallocated responses (\cref{subsec:session_impl}) & 3.55 M/s & 12.6\% \\
Disable 0-copy request processing (\cref{subsubsec:zero_copy_req}) & 3.05 M/s & 14.0\% \\
\end{tabular}
\caption{Impact of disabling optimizations on small RPC rate (CX4)}
\label{tab:commoncase}
\end{center}
\end{table}

Disabling congestion control increases eRPC's request rate on CX4 ($B = 3$)
from \Mrps{4.96} to \Mrps{5.44}. This shows that the overhead of our optimized
congestion control is only 9\%.

\paragraph{Factor analysis.} How important are eRPC's common-case
optimizations? Table~\ref{tab:commoncase} shows the performance impact of
\emph{disabling} some of eRPC's common-case optimizations on CX4; other
optimizations such as our single-DMA msgbuf format and unsignaled transmissions
cannot be disabled easily. For our baseline, we use $B = 3$ and enable
congestion control. Disabling all three congestion control optimizations
(\S~\ref{subsubsec:cc_opt}) reduces throughput to \Mrps{4.3}, increasing the
overhead of congestion control from 9\% to 20\%. Further disabling preallocated
responses and zero-copy request processing reduces throughput to \Mrps{3},
which is 40\% lower than eRPC's peak throughput. \emph{We therefore conclude
that optimizing for the common case is both necessary and sufficient for
high-performance RPCs.}

\subsection{Session scalability}
\label{subsec:session_scalability}
\begin{figure}
	\includegraphics[width=0.45\textwidth]{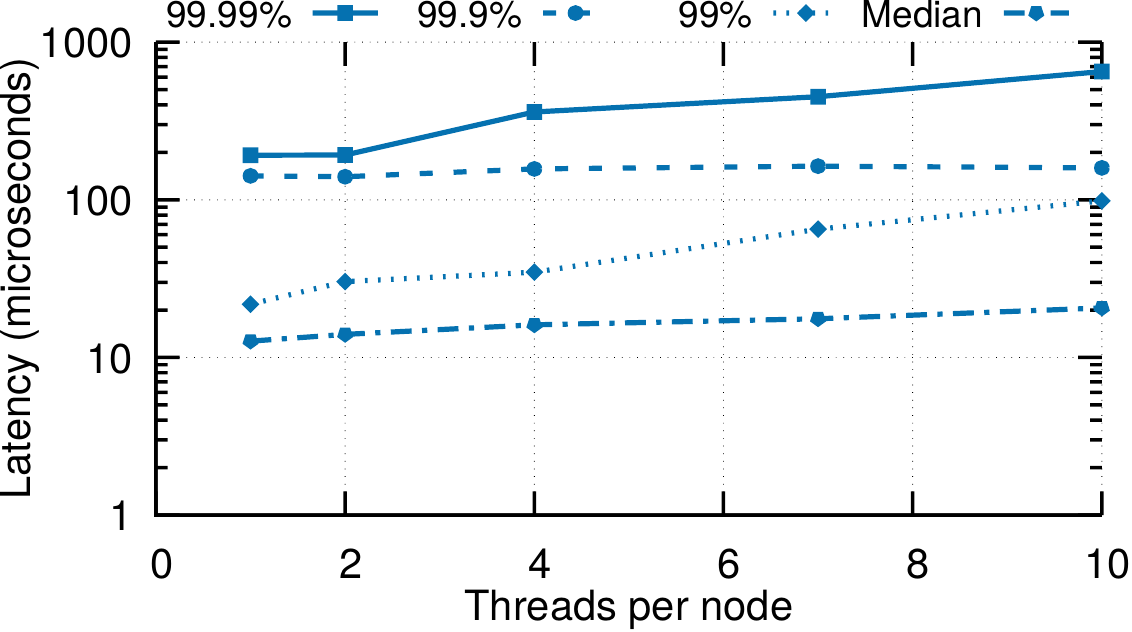}
  \caption{Latency with increasing threads on 100 CX4 nodes}
  \label{fig:cx4_100_latency}
\end{figure}

We evaluate eRPC's scalability on CX4 by increasing the number of nodes in the
previous experiment ($B = 3$) to 100. The five ToR switches in CX4 were
assigned between 14 and 27 nodes each by CloudLab. Next, we increase the number
of threads per node: With $T$ threads per node, there are $100T$ threads in the
cluster; each thread creates a client-mode session to $100T - 1$ threads.
Therefore, each node hosts $T * (100T - 1)$ client-mode sessions, and an equal
number of server-mode sessions. Since CX4 nodes have 10 cores, each node
handles up to 19980 sessions. This is a challenging traffic pattern that
resembles distributed online transaction processing (OLTP) workloads, which
operate on small data items~\cite{Dragojevic:sosp2015, Wei:sosp2015,
Kalia:osdi2016, Zamanian:vldb2017}.

With 10 threads/node, each node achieves \Mrps{12.3} on average. At \Mrps{12.3},
each node sends and receives 24.6 million packets per second (packet
size = \byte{92}), corresponding to \Gbps{18.1}. This is close to the link's
achievable bandwidth (\Gbps{23} out of \Gbps{25}), but is somewhat smaller
because of oversubscription. We observe retransmissions with more than two
threads per node, but the retransmission rate stays below 1700 packets per
second per node. 

Figure~\ref{fig:cx4_100_latency} shows the RPC latency statistics. The
median latency with one thread per node is \us{12.7}. This is higher than the
\us{3.7} for CX4 in Table~\ref{tab:small_rpc_latency} because most RPCs now go
across multiple switches, and each thread keeps 60 RPCs in flight, which adds
processing delay. Even with 10 threads per node, eRPC's 99.99th percentile
latency stays below \us{700}.

These results show that eRPC can achieve high message rate, bandwidth, and
scalability, and low latency in a large cluster with lossy Ethernet.
Distributed OLTP has been a key application for lossless RDMA fabrics; our
results show that it can also perform well on lossy Ethernet.

\subsection{Large RPC bandwidth}
\label{subsec:large_rpc_bandwidth}
\begin{figure}
  \centering
	\includegraphics[width=0.45\textwidth]{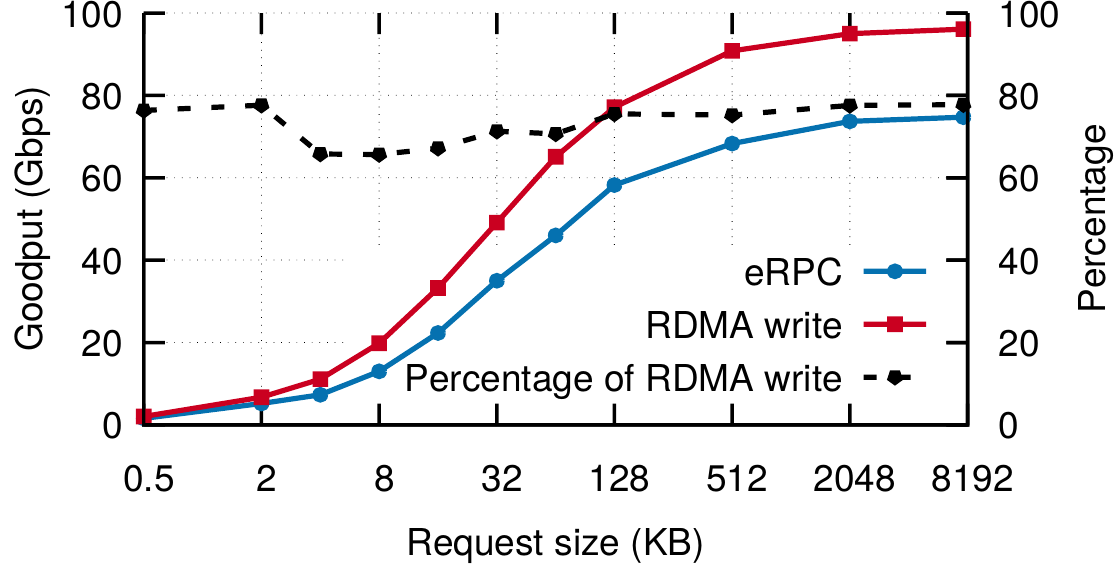}
  \caption{Throughput of large transfers over \Gbps{100} InfiniBand}
	\label{fig:large_rpc_tput}
\end{figure}
We evaluate eRPC's bandwidth using a client thread that sends large messages to
a remote server thread. The client sends $R$-byte requests and keeps one
request outstanding; the server replies with a small \byte{32} response. We use
up to \mbyte{8} requests, which is the largest message size supported by
eRPC\@. We use 32 credits per session. To understand how eRPC performs relative
to hardware limits, we compare against $R$-byte RDMA writes, measured using
\texttt{perftest}.

On the clusters in Table~\ref{tab:clusters}, eRPC gets bottlenecked by network
bandwidth in this experiment setup. To understand eRPC's performance limits, we
connect two nodes in the CX5 cluster to a \Gbps{100} switch via ConnectX-5
InfiniBand NICs. (CX5 is used as a \GbE{40} cluster in the rest of this paper.)
Figure~\ref{fig:large_rpc_tput} shows that eRPC achieves up to \Gbps{75} with
one core. eRPC's throughput is at least 70\% of RDMA write throughput for
\kbyte{32} or larger requests.

In the future, eRPC's bandwidth can be improved by freeing-up CPU cycles.
First, on-die memory copy accelerators can speed up copying data from RX ring
buffers to request or response msgbufs~\cite{www-intel-ioat,
Dalton:nsdi2018}. Commenting out the memory copies at the server increases
eRPC's bandwidth to \Gbps{92}, showing that copying has substantial overhead.
Second, cumulative credit return and request-for-response
(\S~\ref{sec:protocol_messages}) can reduce packet processing overhead.

\begin{table}
\begin{center}
\small
\begin{tabular}{llllll}
\textbf{Loss rate} & $10^{-7}$ & $10^{-6}$ & $10^{-5}$ & $10^{-4}$ & $10^{-3}$ \\
\midrule
\textbf{Bandwidth (Gbps)} & 73 & 71 & 57 & 18 & 2.5 \\
\end{tabular}
\caption{eRPC's 8 MB request throughput with packet loss}
\label{tab:large_rpc_loss}
\end{center}
\end{table}

Table~\ref{tab:large_rpc_loss} shows the throughput with $R$~=~\mbyte{8} (the
largest size supported by eRPC), and varying, artificially-injected packet loss
rates. With the current \ms{5} RTO, eRPC is usable while the loss probability
is up to .01\%, beyond which throughput degrades rapidly. We believe that this
is sufficient to handle packet corruptions. RDMA NICs can handle a somewhat
higher loss rate (.1\%)~\cite{Zhu:sigcomm2015}.

\subsection{Effectiveness of congestion control}
\label{subsec:cc_eval}
\begin{table}
\begin{center}
\small
\begin{tabular}{lSSS@{}}
\textbf{Incast degree} & \textbf{Total bw} & \textbf{50\% RTT} & \textbf{99\% RTT} \\
\midrule
20 & \Gbps{21.8} & \us{39} & \us{67} \\
20 (no cc) & \Gbps{23.1} & \us{202} & \us{204} \\
\midrule
50 & \Gbps{18.4} & \us{34} & \us{174} \\
50 (no cc) & \Gbps{23.0} & \us{524} & \us{524} \\
\midrule
100 & \Gbps{22.8} & \us{349} & \us{969} \\
100 (no cc) & \Gbps{23.0} & \us{1056} & \us{1060} \\
\end{tabular}
\caption{Effectiveness of congestion control (cc) during incast}
\label{tab:cc_incast}
\end{center}
\end{table}

We evaluate if our congestion control is successful at reducing switch
queueing. We create an incast traffic pattern by increasing the number of
client nodes in the previous setup ($R$~=~\mbyte{8}). The one server node acts as the
incast victim. During an incast, queuing primarily
happens at the victim's ToR switch. We use per-packet RTTs measured at the
clients as a proxy for switch queue length~\cite{Mittal:sigcomm2015}.

Table~\ref{tab:cc_incast} shows the total bandwidth achieved by all flows and
per-packet RTT statistics on CX4, for 20, 50, and 100-way incasts (one flow per
client node). We use two configurations: first with eRPC's optimized congestion
control, and second with no congestion control. Disabling our common-case
congestion control optimizations does not substantially affect the RTT
statistics, indicating that these optimizations do not reduce the quality of
congestion control.

Congestion control successfully handles our target workloads of up to 50-way
incasts, reducing median and 99th percentile queuing by over 5x and 3x,
respectively. For 100-way incasts, our implementation reduces median queueing
by 3x, but fails to substantially reduce 99th percentile queueing. This is in
line with~\citet[\S~4.3]{Zhu:conext2016}'s analysis, which shows that
Timely-like protocols work well with up to approximately 40 incast flows.

The combined incast throughput with congestion control is within 20\% of the
achievable \Gbps{23}. We believe that this small gap can be further reduced
with better tuning of Timely's many parameters. Note that we can also support
ECN-based congestion control in eRPC, which may be a better congestion
indicator than RTT~\cite{Zhu:conext2016}.

\paragraph{Incast with background traffic.} Next, we augment the setup above to
mimic an experiment from Timely~\cite[Fig 22]{Mittal:sigcomm2015}: we create
one additional thread at each node that is not the incast victim. These threads
exchange latency-sensitive RPCs (\kbyte{64} request and response), keeping one
RPC outstanding. During a 100-way incast, the 99th percentile latency of these
RPCs is \us{274}. This is similar to Timely's latency ($\approx$200-\us{300})
with a 40-way incast over a \GbE{20} lossless RDMA fabric. Although the two
results cannot be directly compared, this experiment shows that the latency achievable with
software-only networking in commodity, lossy datacenters is comparable to
lossless RDMA fabrics, even with challenging traffic patterns.

\section{Full-system benchmarks}
\label{sec:full_system}
In this section, we evaluate whether eRPC can be used in real applications with
unmodified existing storage software: We build a state machine replication
system using an open-source implementation of Raft~\cite{Ongaro:2014}, and a
networked ordered key-value store using Masstree~\cite{Mao2012:eurosys2012}.

\subsection{Raft over eRPC}
\label{subsec:raft}
State machine replication (SMR) is used to build fault-tolerant services. An
SMR service consists of a group of server nodes that receive commands from
clients. SMR protocols ensure that each server executes the same sequence of
commands, and that the service remains available if servers fail.
Raft~\cite{Ongaro:2014} is such a protocol that takes a \emph{leader}-based
approach: Absent failures, the Raft replicas have a stable leader to
which clients send commands; if the leader fails, the remaining Raft servers
elect a new one. The leader appends the command to replicas' logs, and it
replies to the client after receiving acks from a majority of
replicas.

SMR is difficult to design and implement correctly~\cite{Hawblitzel:sosp2015}:
the protocol must have a specification and a proof (e.g., in TLA+), and the
implementation must adhere to the specification. We avoid this difficulty by
using an existing implementation of Raft~\cite{www-willemt-raft}. (It had no
distinct name, so we term it LibRaft.)  We did not write LibRaft ourselves; we
found it on GitHub and used it as-is.  LibRaft is well-tested with fuzzing over
a network simulator and 150+ unit tests.  Its only requirement is that the user
provide callbacks for sending and handling RPCs---which we implement using
eRPC\@. Porting to eRPC required no changes to LibRaft's code.

We compare against recent consistent replication systems that are built from
scratch for two specialized hardware types. First, NetChain~\cite{Jin:nsdi2018}
implements chain replication over programmable switches. Other replication
protocols such as conventional primary-backup and Raft are too complex to
implement over programmable switches~\cite{Jin:nsdi2018}. Therefore, despite
the protocol-level differences between LibRaft-over-eRPC and NetChain, our
comparison helps understand the relative performance of end-to-end CPU-based
designs and switch-based designs for in-memory replication. Second,
Consensus~in~a~Box~\cite{Istvan:nsdi2016} (called ZabFPGA here), implements
ZooKeeper's atomic broadcast protocol~\cite{Hunt:zookeeper:2010} on FPGAs. eRPC
also outperforms DARE~\cite{Poke:hpdc2015}, which implements SMR over RDMA; we
omit the results for brevity.

\paragraph{Workloads.} We mimic NetChain and ZabFPGA's experiment setups for
latency measurement: we implement a 3-way replicated in-memory key-value store,
and use one client to issue PUT requests. The replicas' command logs and
key-value store are stored in DRAM\@. NetChain and ZabFPGA use \byte{16} keys,
and 16--\byte{64} values; we use \byte{16} keys and \byte{64} values. The
client chooses PUT keys uniformly at random from one million keys. While
NetChain and ZabFPGA also implement their key-value stores from scratch, we reuse
existing code from MICA~\cite{Li:isca2015}. We compare eRPC's performance on
CX5 against their published numbers because we do not have the hardware to run
NetChain or ZabFPGA\@. Table~\ref{tab:raft_latency} compares the latencies of the
three systems.

\begin{table}
\small
\centering
\begin{tabular}{llSS@{}}
\textbf{Measurement} & \textbf{System} & \textbf{Median} & \textbf{99\%} \\
\midrule
\multirow{2}{*}{\shortstack[l]{Measured at client}} & NetChain & \us{9.7} & N/A \\
& eRPC & \us{5.5} & \us{6.3} \\
\midrule
\multirow{2}{*}{\shortstack[l]{Measured at leader}} & ZabFPGA & \us{3.0} & \us{3.0} \\
& eRPC & \us{3.1} & \us{3.4} \\
\end{tabular}
\caption{Latency comparison for replicated PUTs}
\label{tab:raft_latency}
\end{table}

\subsubsection{Comparison with NetChain}
NetChain's key assumption is that software networking adds 1--2 orders of
magnitude more latency than switches~\cite{Jin:nsdi2018}.
However, we have shown that eRPC adds \ns{850}, which is only around 2x higher
than latency added by current programmable switches
(\ns{400}~\cite{www-aurora-tofino}).

Raft's latency over eRPC is \us{5.5}, which is substantially lower than
NetChain's \us{9.7}. This result must be taken with a grain of salt: On the one
hand, NetChain uses NICs that have higher latency than CX5's NICs.
On the other hand, it has numerous limitations, including
key-value size and capacity constraints, serial chain replication whose latency
increases linearly with the number of replicas, absence of congestion control,
and reliance on a complex and external failure detector. The main takeaway is
that microsecond-scale consistent replication is achievable in commodity
Ethernet datacenters with a general-purpose networking library.

\subsubsection{Comparison with ZabFPGA}
Although ZabFPGA's SMR servers are FPGAs, the clients are commodity
workstations that communicate with the FPGAs over slow kernel-based TCP\@. For
a challenging comparison, we compare against ZabFPGA's commit latency measured
at the leader, which involves only FPGAs. In addition, we consider its ``direct
connect'' mode, where FPGAs communicate over point-to-point links (i.e.,
without a switch) via a custom protocol. Even so, eRPC's median leader commit
latency is only 3\% worse.

An advantage of specialized, dedicated hardware is low jitter. This is
highlighted by ZabFPGA's negligible leader latency variance. This advantage does
not carry over directly to end-to-end latency~\cite{Istvan:nsdi2016} because
storage systems built with specialized hardware are eventually accessed by clients
running on commodity workstations.

\subsection{Masstree over eRPC}
Masstree~\cite{Mao2012:eurosys2012} is an ordered in-memory key-value store. We
use it to implement a single-node database index that supports low-latency
point queries in the presence of less performance-critical longer-running
scans. This requires running scans in worker threads. We use CX3 for
this experiment to show that eRPC works well on InfiniBand.

We populate a Masstree server on CX3 with one million random \byte{8} keys
mapped to \byte{8} values. The server has 16 Hyper-Threads, which we divide
between 14 dispatch threads and 2 worker threads. We run 64 client threads
spread over 8 client nodes to generate the workload. The workload consists of
99\% GET(key) requests that fetch a key-value item, and 1\% SCAN(key) requests
that sum up the values of 128 keys succeeding the key. Keys are chosen
uniformly at random from the inserted keys. Two outstanding requests per client
was sufficient to saturate our server.

We achieve 14.3 million GETs/s on CX3, with \us{12} 99th percentile GET latency.
If the server is configured to run only dispatch threads, the 99th percentile
GET latency rises to \us{26}. eRPC's median GET latency under low load is
\us{2.7}. This is around 10x faster than Cell's single-node B-Tree that uses
multiple RDMA reads~\cite{Mitchell:usenix2016}. Despite Cell's larger key/value
sizes (\byte{64}/\byte{256}), the latency differences are mostly from RTTs:  At
\Gbps{40}, an additional \byte{248} takes only \ns{50} more time to transmit.

\section{Related work}
\label{sec:related}

\paragraph{RPCs.} There is a vast amount of literature on RPCs. The practice of
optimizing an RPC wire protocol for small RPCs originates with
\citet{Birrell:1984}, who introduce the idea of an implicit-ACK\@. Similar to
eRPC, the Sprite RPC system~\cite{Welch:1986} directly uses raw datagrams and
performs retransmissions only at clients. The Direct Access File
System~\cite{DeBergalis:2003} was one of the first to use RDMA in RPCs. It uses
SEND/RECV messaging over a connected transport to initiate an RPC, and RDMA
reads or writes to transfer the bulk of large RPC messages. This design is
widely used in other systems such as NFS's RPCs~\cite{Callaghan:2003} and some
MPI implementations~\cite{Liu:mpi2004}.
In eRPC, we chose to transfer all data over datagram messaging to avoid
the scalability limits of RDMA\@. Other RPC systems that use RDMA include
Mellanox's Accelio~\cite{www-mellanox-accelio} and RFP~\cite{su:eurosys2017}.
These systems perform comparably to FaRM's RPCs, which are slower than eRPC at
scale by an order of magnitude.

\paragraph{Co-design.} There is a rapidly-growing list of projects that
co-design distributed systems with the network. This includes key-value
stores~\cite{Mitchell:usenix2013, Lim:nsdi2014, Kalia:sigcomm2014,
Wang:sc2015}, distributed databases and transaction processing
systems~\cite{Dragojevic:nsdi2014, Wei:sosp2015, Chen:eurosys2016,
Zamanian:vldb2017}, state machine replication~\cite{Poke:hpdc2015,
Istvan:nsdi2016}, and graph-processing systems~\cite{Shi:osdi2016}. We believe
the availability of eRPC will motivate researchers to investigate how much
performance these systems can achieve without sacrificing the networking
abstraction. On the other hand, there is a smaller set of recent projects that
also prefer RPCs over co-design, including RAMCloud, FaSST, and the distributed
data shuffler by \citet{Liu:eurosys2017}. However, their RPCs lack either
performance (RAMCloud) or generality (FaSST), whereas eRPC provides both.

\section{Conclusion}
\label{sec:concl}

eRPC is a fast, general-purpose RPC system that provides an attractive
alternative to putting more functions in network hardware, and specialized
system designs that depend on these functions. eRPC's speed comes from
prioritizing common-case performance, carefully combining a wide range of old
and new optimizations, and the observation that switch buffer capacity far
exceeds datacenter BDP. eRPC delivers performance that was until now believed
possible only with lossless RDMA fabrics or specialized network hardware. It
allows unmodified applications to perform close to the hardware limits. Our
ported versions of LibRaft and Masstree are, to our knowledge, the fastest
replicated key-value store and networked database index in the academic
literature, while operating end-to-end without additional network support.

\vspace{10pt}
\noindent
\textbf{Acknowledgments}
\small{
We received valueable insights from our SIGCOMM and NSDI reviewers, Miguel
Castro, John Ousterhout, and Yibo Zhu. Sol Boucher, Jack Kosaian, and Hyeontaek
Lim helped improve the writing. CloudLab~\cite{Ricci:login14} and
Emulab~\cite{White+:osdi2002} resources were used in our experiments. This work
was supported by funding from the National Science Foundation under awards
CCF-1535821 and CNS-1700521, and by Intel via the Intel Science and Technology
Center for Visual Cloud Systems (ISTC-VCS)\@. Anuj Kalia is supported by the
Facebook Fellowship.}
\label{lastpage}

\clearpage

\normalsize
\titleformat{\section}%
  {\bf}{Appendix~\thesection.\quad}{0pt}{}
\appendix
\section{\lowercase{e}RPC's NIC memory footprint}
\label{sec:erpc_nic_footprint}
Primarily, four on-NIC structures contribute to eRPC's NIC memory footprint:
the TX and RX queues, and their corresponding completion queues. The TX queue
must allow sufficient pipelining to hide PCIe latency; we found that 64 entries
are sufficient in all cases. eRPC's TX queue and TX completion queue have 64
entries by default, so their footprint does not depend on cluster size. The
footprint of on-NIC page table entries required for eRPC is negligible because
we use \mbyte{2} hugepages~\cite{Dragojevic:nsdi2014}.

As discussed in Section~\ref{subsubsec:session_credits}, eRPC's RQs must have
sufficient descriptors for all connected sessions. If traditional RQs are used,
their footprint grows with the number of connected sessions supported. Modern
NICs (e.g., ConnectX-4 and newer NICs from Mellanox)
support \emph{multi-packet} RQ descriptors that
specify multiple contiguous packet buffers using base address, buffer size, and
number of buffers. With eRPC's default configuration of 512-way RQ descriptors,
RQ size is reduced by 512x, making it negligible. This optimization has the
added advantage of almost eliminating RX descriptor DMA, which is now needed only
once every 512 packets. While multi-packet RQs were originally designed for
large receive offload of one message~\cite{www-mellanox-ofed-release-notes}, we
use this feature to receive packets of independent messages. 

What about the RX completion queue (CQ)? By default, NICs expect the RX CQ to
have sufficient space for each received packet, so using multi-packet RQ
descriptors does not reduce CQ size. However, eRPC does not need the
information that the NIC DMA-writes to the RX CQ entries. It needs only the number
of new packets received. Therefore, we shrink the CQ by
allowing it to \emph{overrun}, i.e., we allow the NIC to overwrite existing
entries in the CQ in a round-robin fashion. We poll the overrunning CQ to check
for received packets. It is possible to use a RX CQ with only one entry, but we
found that doing so causes cache line contention between eRPC's threads and
the CPU's on-die PCIe controller. We solve this issue by using 8-entry CQs,
which makes the contention negligible.

\section{Handling node failures}
\label{sec:machine_failure}
eRPC launches a session management thread that handles sockets-based management
messaging for creating and destroying sessions, and detects failure of remote
nodes with timeouts. When the management thread suspects a remote node failure,
each dispatch thread with sessions to the remote node acts as follows. First,
it flushes the TX DMA queue to release msgbuf references held by the NIC.
For client sessions, it waits for the rate limiter to transmit any queued
packets for the session, and then invokes continuations for pending requests
with an error code. For server-mode sessions, it frees session resources
after waiting (non-blocking) for request handlers that have not enqueued a
response.

\section{Rate limiting with zero-copy}
\label{sec:rl_zero_copy}
Recall the request retransmission example discussed in
\S~\ref{subsubsec:buffer_ownership}: On receiving the response for the first
copy of a retransmitted request, we wish to ensure that the rate limiter does
not contain a reference to the retransmitted copy. Unlike eRPC's NIC DMA queue
that holds only a few tens of packets, the rate limiter tracks up to
milliseconds worth of transmissions during congestion. As a result, flushing it
like the DMA queue is too slow. Deleting references from the rate limiter
turned out to be too complex: Carousel requires a bounded difference between
the current time and a packet's scheduled transmission time for correctness, so
deletions require rolling back Timely's internal rate computation state. Each
Timely instance is shared by all slots in a session
(\S~\ref{subsec:session_impl}), which complicates rollback.

We solve this problem by dropping response packets received while a
retransmitted request is in the rate limiter. Each such response indicates a
false positive in our retransmission mechanism, so they are rare. This solution
does not work for the NIC DMA queue: since we use unsignaled transmission, it
is generally impossible for software to know whether a request is in the DMA
queue without flushing it.


\setlength{\bibsep}{2pt}
\bibliography{ref}
\bibliographystyle{abbrvnat}

}{
}

\end{document}